\documentclass[12pt]{article}
 \pdfoutput=1
\usepackage{amsfonts,amsthm,amsmath,amssymb,slashed}
\usepackage[textwidth = 500 pt, textheight = 630 pt]{geometry}
\usepackage{latexsym,amsfonts,amsmath,amssymb,theorem,dsfont,wasysym}
\usepackage{amssymb}
\usepackage{amsmath}
\usepackage{amstext}
\usepackage{graphicx,epsfig}
\usepackage{epsfig}
\usepackage{verbatim} 
\usepackage{fancyhdr}
\usepackage{fancybox}
\usepackage{color}
\usepackage{ulem,bbold}
\usepackage{enumitem}
\usepackage{subfigure}
\usepackage{bbm}
\usepackage{parskip}
\usepackage[numbers,sort&compress]{natbib}

\definecolor{MyDarkBlue}{rgb}{0.15,0.15,0.45}
\usepackage[linktocpage=true]{hyperref}
\hypersetup{
colorlinks=true,
citecolor=MyDarkBlue,
linkcolor=MyDarkBlue,
urlcolor=MyDarkBlue,
}

\usepackage[numbers,sort&compress]{natbib}

\def\beq{\begin{eqnarray}}
\def\eeq{\end{eqnarray}}

\def\({\left(}
\def\){\right)}

\def\mpl{M_{\rm pl}}

\def\f{\varphi}

\newcommand{\be}{\begin{equation}}
\newcommand{\ee}{\end{equation}}
\newcommand{\la}{\langle}
\newcommand{\ra}{\rangle}
\def\ea{\end{eqnarray}}
\def\ba{\begin{eqnarray}}

\def\beq{\begin{eqnarray}}
\def\eeq{\end{eqnarray}}

\def\({\left(}
\def\){\right)}

\def\mpl{M_{\rm pl}}

\def\p{\partial}
\def\la{\langle}
\def\ra{\rangle}

\def\lsim{\mathrel{\rlap{\lower3pt\hbox{\hskip0pt$\sim$}}
     \raise1pt\hbox{$<$}}}         
\def\gsim{\mathrel{\rlap{\lower4pt\hbox{\hskip1pt$\sim$}}
     \raise1pt\hbox{$>$}}}         

\def\lsim{\mathrel{\rlap{\lower3pt\hbox{\hskip0pt$\sim$}}
     \raise1pt\hbox{$<$}}}         
\def\gsim{\mathrel{\rlap{\lower4pt\hbox{\hskip1pt$\sim$}}
     \raise1pt\hbox{$>$}}}         

\usepackage{amsmath}
\usepackage{amsfonts}
\usepackage{verbatim}
\usepackage{graphicx}
\usepackage{subfigure}
\usepackage{amssymb}
\usepackage[T1]{fontenc}
\usepackage{slashed}

\begin{document}

\renewcommand{\thefootnote}{\fnsymbol{footnote}}

\makeatletter
\@addtoreset{equation}{section}
\makeatother
\renewcommand{\theequation}{\thesection.\arabic{equation}}

\rightline{}
\rightline{}



\begin{center}
{\Large \bf{On Corpuscular Theory of Inflation}}

 \vspace{1truecm}
\thispagestyle{empty} \centerline{\large  {Lasha  Berezhiani}
}

 \textit{Department of Physics, Princeton University, \\
Jadwin Hall, Princeton, NJ 08540, USA
  }

\end{center}  
 
\begin{abstract}

In order to go beyond the mean-field approximation, commonly used in the inflationary computations, an identification of the quantum constituents of the inflationary background is made. In particular, the homogeneous scalar field configuration is represented as a Bose-Einstein condensate of the off-shell inflaton degrees of freedom, with mass significantly screened by the gravitational binding energy.  The gravitational counterpart of the classical background is considered to be a degenerate state of the off-shell longitudinal gravitons with the frequency of the order of the Hubble scale. As a result, the origin of the density perturbations in the slow-roll regime is identified as an uncertainty in the position of the constituent inflatons. While in the regime of eternal inflation, the scattering of the constituent gravitons becomes the relevant source of the density perturbations. The gravitational waves, on the other hand, originate from the annihilation of the constituent longitudinal gravitons at all energy scales. This results in the quantum depletion of the classical background. Leading to the upper bound on the number of e-folds, after which the semiclassical description is expected to break down, which is estimated to be of the order of the entropy of the initial Hubble patch.

\end{abstract}

\newpage
\setcounter{page}{1}

\renewcommand{\thefootnote}{\arabic{footnote}}
\setcounter{footnote}{0}

\linespread{1.1}
\parskip 4pt

\section{Introduction}

The inflationary paradigm is a tantalizing idea of producing the entire observable Universe from a small homogeneous causally connected patch as a result of the accelerated expansion \cite{Guth:1980zm,Linde:1981mu,Albrecht:1982wi}.

The inflationary computations are usually done in the mean-field approximation, in which one considers the quantum fluctuations around a fixed classical background \cite{mukhanov,starobinsky,hawking,guth,bardeen} (see also \cite{Maldacena:2002vr} and references therein). Although this approach has been proven to be very powerful, there may be some quantum effects that lie beyond the reach of this semi-classical approximation. Especially if we would like to understand the nature of slow-roll eternal inflation \cite{Steinhardt:1982kg,Linde:1982ur,Vilenkin:1983xq,Linde:1986fd,Linde:1993xx,Creminelli:2008es}, which is governed by quantum dynamics.

Recently, a microscopic description of inflation has been proposed in \cite{Dvali:2013eja}. In which, the classical background is considered to be composed of quanta, just like any other classical object. This approach is similar to the particle number conserving formalism for Bose-Einstein condensates used in condensed matter literature. Even though in case of inflation the particle number is not expected to be precisely conserved, this formalism will safeguard us from an unphysical particle production which may be an artifact of the mean field approximation.

The main idea of \cite{Dvali:2013eja} was to think of the classical backgrounds to be built on top of the Minkowski space using certain creation operators. The reason for choosing the Minkowski space as a base state for the construction is twofold. Firstly, it is a true vacuum state for the conventional models of inflation\footnote{For simplicity, we will be assuming the absence of the cosmological constant throughout this work.}. Secondly, there is no particle production on the Minkowski space, therefore in the picture of \cite{Dvali:2013eja} the particle production during inflation should be understood as a result of the dynamics of the constituent degrees of freedom, rather than as the vacuum process.  

In order to motivate the quantum picture further, let us consider the history of our Universe. In particular, let us begin by radiation dominated era. In cosmological computations we treat radiation as a classical fluid with relativistic equation of state. However, this is only an approximation and in reality this fluid is composed of photons; which simply means that if we start removing photons from a given region of the universe, eventually we would be left with a Minkowski space. This immediately points towards the quantum compositeness of the classical metric itself; as the microscopic modification to the photon gas, must lead to the microscopic modification to the gravitational background.

Going further back into the past, and assuming that the hot big bang was preceded by inflation, we reach the reheating era. At this time it is a common practice to treat the oscillating homogeneous scalar field as a Bose-Einstein condensate (BEC) of inflatons. After all, in order to finish in the quantum state with finite density of quantum constituents, we need to begin with such a state. 

Therefore, it is natural to describe the scalar background as some sort of BEC even past the reheating point. During inflation the scalar field undergoes a slow-roll instead of the rapid oscillations, which suggests that the inflationary BEC may need to differ from the reheating BEC.

We will indeed show that the inflationary background can be described as a degenerate state of off-shell inflatons with significantly screened mass. The mass of the constituents is in fact the only distinguishing factor between the two condensates. Moreover, we will give a microscopic account of the well-known effects such as Gibbons-Hawking radiation \cite{Gibbons:1977mu}, as well as scalar \cite{mukhanov} and tensor \cite{Starobinsky:1979ty} modes of inflationary perturbations.

\section{Classical Picture}

For simplicity we will consider the model in which inflation is driven by the single massive scalar field, without self-interactions, minimally coupled to gravity \cite{Linde:1983gd}
\beq
S=\int {\rm d} ^4x\sqrt{-g}\left( \mpl^2 R-\frac{1}{2}g^{\mu\nu}\p_\mu\f\p_\nu\f - \frac{1}{2}m^2 \f^2\right)\,.
\label{action}
\eeq
The classical dynamics of the homogeneous background follows from the equation of motion for the scalar field and the Friedmann's equation
\beq
\label{inflatoneq}
&&\ddot{\f}+3H\dot{\f}+m^2\f=0\,,\\
&&\mpl^2H^2=\frac{1}{2}\left( \dot{\f}^2 +m^2\f^2\right)\,.
\label{friedmann}
\eeq
Depending on the value of the slow-roll parameter $\epsilon\equiv m^2/H^2$, there are two distinct regimes of the evolution. In particular, it is easy to see that for $\epsilon\ll 1$, the equations of motion simplify in the absence of the initial kinetic energy and reduce to
\beq
&&\dot{\f}\simeq -\frac{m^2}{3H}\f\,,\\
&&\mpl^2H^2\simeq\frac{1}{2}m^2\f^2\,.
\eeq
In this approximation, the scalar field is slowly rolling down the potential. In fact, it does not change significantly within the Hubble time, resulting in the nearly constant $H$ and consequently in the quasi-de Sitter universe.

For $\epsilon\gg 1$, on the other hand, the scalar field undergoes damped oscillations with frequency $m$ and hence behaves like a degenerate gas of dust. In the presence of other light species in the spectrum, the universe would reheat at this point.

\section{Quantum Picture}

Following \cite{Dvali:2013eja}, We would like to begin the discussion from the reheating era. As we have already mentioned, for $\epsilon\gg 1$ the scalar field behaves like a dust. Therefore, it should be described as a BEC of nearly on-shell $\f$-quanta in $k=0$ state. Indeed, during reheating it is a common practice to treat the inflaton background as a degenerate quantum state.

Because of this, it is natural to try to describe the scalar background as some sort of BEC even during inflation. Since the inflationary background is a homogeneous field configuration, it should be viewed as a condensate of the finite number of $\f$-quanta in $k=0$ state. Other properties of this condensate can be identified by matching to the known semi-classical results. For instance, we need to obtain the correct amplitude of density perturbations.

There are two potential sources of the density perturbations: the uncertainty principle and the scattering of the constituents.

The scalar constituents have vanishing wavenumber. Therefore, their wavefunction is completely delocalized throughout the entire universe;  which means that the number of $\varphi$-quanta within a given Hubble patch is not fixed. In fact, it must undergo quantum fluctuations due to the uncertainty principle. Ignoring interactions, one can show that the number of quanta in a given region of the universe is given by the Poisson's distribution\footnote{Strictly speaking, this statement depends on the quantum state of inflatons. We will elaborate on this in the appendix.}. This means that if the expected number of scalar quanta in a given region, let's say the Hubble patch, is $N_\varphi$ then the typical number fluctuation is $\delta N_\varphi=\sqrt{N_\varphi}$.

In order to estimate the amplitude of perturbations we need to be more specific about the scalar constituents. During the reheating era, inflaton is undergoing the damped oscillations. Therefore, at this time the scalar background is a standard BEC of $\varphi$-quanta with $k=0$ and the energy $\omega=m$. The question is what kind of condensate does the inflationary stage correspond to.

Let us assume, for now, that during inflation the scalar background can still be thought of as a collection of $k=0$ and $\omega=m$ weakly-interacting quanta. Then, in order to source the curvature $H$ we need the following number of quanta within the Hubble patch
\beq
N_\varphi=\frac{\mpl^2}{mH}\,.
\label{number1}
\eeq
The typical density perturbation due to the uncertainty principle would be
\beq
\label{den1}
\delta\rho=\sqrt{N_\varphi}mH^3\,.
\eeq
In order to estimate the corresponding $\delta\varphi$, let us note that
\beq
\delta\rho=m^2\varphi\delta\varphi=mH\mpl\delta\varphi\,.
\label{scalarpert}
\eeq
Equating this with \eqref{den1}, we obtain
\beq
\delta\varphi=H\sqrt{\frac{H}{m}}\,.
\eeq
Which is inconsistent with the standard result $\delta\varphi=H$.

In order to identify the problem in the above computation, it is useful to pause and think about the background we are describing. During reheating the background is undergoing oscillations with the frequency $m$. Hence, when thinking of it as a coherent state, it is natural for the constituents to have $\omega=m$. When it comes to the inflationary stage, on the other hand, the scalar field is undergoing a slow-roll with the rate much smaller than $m$
\beq
\dot{\varphi}=-\frac{m^2}{H}\varphi\,.
\eeq
Therefore, it appears to be more natural to think of the inflaton background as a condensate of the off-shell $\varphi$-quanta in $k=0$ state with
\beq
\omega=m_{eff}\equiv\frac{m^2}{H}\,.
\eeq
Since $m_{eff}\ll m$ we will need more degrees of freedom than \eqref{number1} in order to source the curvature $H$. In particular, we have
\beq
N_\varphi=\frac{H^2\mpl^2}{m_{eff} H^3}=\frac{\mpl^2}{m^2}\,.
\label{number2}
\eeq
Now, the typical density perturbation is given by
\beq
\delta\rho=\sqrt{N_\varphi}m_{eff}H^3=\mpl m H^2\,,
\label{UPperturb}
\eeq
which corresponds to the correct amplitude of the scalar fluctuation on the horizon scale; equating this with \eqref{scalarpert} gives us
\beq
\delta \varphi=H\,.
\eeq
Another source of scalar perturbations is the scattering of the constituents, which we will discuss shortly. According to \cite{Dvali:2013eja}, the background geometry of inflation quantum mechanically corresponds to the degenerate state of the off-shell longitudinal gravitons with $\omega=H$ and
\beq
N=\frac{\mpl^2}{H^2}\,,
\eeq
number of them per Hubble patch.
Therefore, we are describing the interior of the horizon by the following number eigenstate\footnote{This state ought to reproduce the classical background in large-$N$ limit, which means that it could have been chosen to be a coherent state. However, for simplicity we will treat it as a number eigenstate.
}
\beq
\left|N^{\omega=H},N_\varphi^{\omega=\frac{m^2}{H}}\right\rangle\,, \qquad \text{with} \qquad N=\frac{\mpl^2}{H^2},~~N_\varphi=\frac{\mpl^2}{m^2}\,.
\eeq

We would like to stress that, the constituents of inflation are not the same as the asymptotic degrees of freedom. For the scalar field, the latter would be created by the usual creation operator $a_k^\dagger$, which creates a $\f$-quantum with the dispersion relation $\omega_k^2=m^2+k^2$. If we were interested in describing a state, in which degrees of freedom behaved as a weakly interacting gas of these massive particles, then we could have considered a state built using these asymptotic creation operators. For instance, the scalar background during the reheating era can be described as such a state. In the regime of the slow-roll inflation, on the other hand, this is not true anymore, simply because the scalar field background does not evolve appropriately. Instead, we need to define a new set of creation operators $b_k^\dagger$, which create quanta with the mass gap $\omega_{k=0}=\frac{m^2}{H}$. The momentum dependence of the dispersion relation is closely related to the dispersion relation of the density perturbations computed using the mean-field approximation. Moreover, in general we should not expect their interaction vertices to be identical to the ones given by \eqref{action}. These points along with other related issues will be discussed in detail in \cite{LB}.

In the current work, we will simply rely on the universality of the gravitational interactions and assume that the interaction vertices are very similar to the ones of the asymptotic theory.
Next, we would like to estimate the interaction rate between the constituents; since we should expect interactions to excite particles out of the condensate state. One of the relevant channels responsible for scalar excitations is
\beq
g+\varphi \longrightarrow g+\varphi\,.
\label{process}
\eeq
This channel is expected to be significant because of the large multiplicity of $\f$-quanta.

Keeping in mind the comments given above, the interaction vertices responsible for this scattering process are schematically given by
\beq
\frac{h}{\mpl}(\partial \f)^2\,,~~\frac{h^2}{\mpl^2}(\partial \f)^2\,,~\ldots\,,
\eeq
where $\f$ and $h$  denote the off-shell inflaton and graviton degrees of freedom.

The rate of process \eqref{process} can be estimated as
\beq
\Gamma=\left(\frac{m_{eff}H}{\mpl^2}\right)^2\frac{H^2}{m_{eff}}NN_\varphi=H\,,
\label{rate}
\eeq
where the expression in parentheses is the $2\rightarrow 2$ scattering amplitude, $H^2/m_{eff}$ represents the kinematic factor and $NN_\varphi$ is the combinatoric factor.\footnote{I am thankful to Jean-Luc Lehners for pointing out an error in the last step of the original \eqref{rate}.}
Therefore, the number of excited quanta within Hubble time is
\beq
\delta N=\delta N_{\varphi}=1\,.
\eeq
The corresponding density perturbation can be estimated as
\beq
\delta \rho=\delta N \cdot H\cdot H^3=H^4\,.
\eeq
Another interaction channel which could result in the significant scalar amplitude is
\beq
g+g\longrightarrow \f + \f\,.
\label{ggff}
\eeq
This corresponds to the annihilation of the constituent gravitons into the pair of inflatons. The simple estimate shows that the rate of this process is also of order $H$, therefore we should expect to produce approximately few $\f$-quanta with energy $H$ through these channels. The estimate for the corresponding magnitude of the density perturbation is given by
\beq
\delta\rho=H^4\,.
\eeq
Comparing this to \eqref{UPperturb}, we find 
\beq
\frac{\delta\rho \Big|_{\rm {uncertainty ~ principle}}}{\delta\rho\Big|_{\rm scattering}}=\frac{\mpl m}{H^2}\,.
\eeq
Therefore, we can see the emergence of the crossover curvature scale
\beq
H_*\equiv \sqrt{\mpl m}\,.
\label{crossover}
\eeq
In the low curvature regime $H<H_*$, the dominant source of the density perturbations is the uncertainty principle. For $H>H_*$, on the other hand, the Hamiltonian process \eqref{ggff} begins to dominate.

For completeness, let us estimate the amplitude of the scalar fluctuations for these two sources separately. Let us begin with the uncertainty principle, for which we have \eqref{UPperturb}. In order to find the scalar amplitude, we need to relate the density perturbation to $\delta\f$. Notice, that \eqref{scalarpert} is true as long as the nonlinear terms in $\delta\f$ are negligible. In order to check the validity of this approximation, we retain the largest nonlinear correction. As a result, \eqref{scalarpert} becomes
\beq
\delta\rho=m^2\f \delta\f+\delta\dot{\f}^2\,.
\eeq
Applying this to the perturbations with frequency of order $H$, we get
\beq
\delta\rho=mH\mpl\delta\f+H^2\delta\f^2\,.
\label{densitypert2}
\eeq
After equating this expression with \eqref{UPperturb} we find that the nonlinear term becomes important for $H>H_*$, resulting in
\beq
\delta\f\Big|_{\rm {uncertainty ~ principle}}=\begin{cases} H\,,  \quad ~\text{for}\quad H \ll H_*\,, \\ H_*\,,  \quad \text{for}\quad H\gg H_*\,.\end{cases}
\eeq
Repeating the computation for the Hamiltonian process \eqref{ggff}, we arrive at
\beq
\delta\f\Big|_{\rm scattering}=\begin{cases} \frac{H^3}{H_*^2}\,,  \quad ~\text{for}\quad H \ll H_*\,, \\ H\,,  \quad ~~\text{for}\quad H\gg H_*\,.\end{cases}
\eeq
To summarize, we find that the scalar amplitude $\delta\f$ is of the order of the curvature scale $H$, irrespective of the energy density. As for the origin of the perturbation, the uncertainty principle is the dominant source of the density perturbations for $H<H_*$, while for $H>H_*$ the scattering of the constituents takes over.

Interestingly, the crossover curvature scale $H_*$ is also the scale above which the universe enters the self-reproductive regime. In other words, at this scale the amplitude of quantum fluctuations begins to exceed the classical displacement of the scalar field within the Hubble time.

Also, it should be noted that \eqref{rate} is finite in $\epsilon \rightarrow 0$ limit; unlike the result of \cite{Dvali:2013eja}, where the corresponding scattering rate was found to diverge in this limit. This means that, in out picture, we can view the de Sitter space as a limiting case of inflation. 

Having identified the source of the density perturbations, let us discuss the production of other particles; those not being the constituents of the background. For instance, the gravitational waves, or any other spectator field, have vanishing occupation number on the background state. This means that the only possible source for their production can be the annihilation of the constituent degrees of freedom into these states. According to \cite{Dvali:2013eja}, the dominant channel for the production of the gravitational waves is
\beq
g+g \longrightarrow g+g\,,\nonumber
\eeq
with the rate of this process being given by
\beq
\Gamma=H\,.
\eeq

It is easy to show that this production rate corresponds to the correct power-spectrum for the tensor modes
\beq
\delta_T=\frac{H}{\mpl}\,.
\label{gg}
\eeq
In the next section, we show that the process responsible for the production of the gravitational waves also gives rise to a quantum clock.

Let us conclude this section by pointing out a caveat. In light of the discussion in the first paragraph of page 5, that the dispersion relations for the constituents can be identified by studying the collective excitations around the classical background, there can be only one dynamical scalar. In other words, both scalar field  and metric backgrounds can be thought of as a collection of the off-shell particles with gap set by the frequency of the classical background, however only one species of the constituents will have different $k$-levels excitable. In particular, if we choose a spatially flat gauge, in which the scalar perturbation of the gauge field is fixed to zero, then we have chosen a formalism with gauge constituents frozen in the ground state; with their dispersion relation being given by $\omega_k=H$, for all $k$. The aftermath of this stipulation would be the prohibition of the process \eqref{process}, the details will be discussed in \cite{LB}. However, by no means should the number of constituent gravitons be considered as fixed; e.g. they could annihilate, resulting in the depletion of the condensate.

\section{Depletion of the Condensate}

As we have discussed in the previous section, the production of the gravitational waves is the result of the annihilation of the constituent longitudinal gravitons. Therefore, this process must drain the graviton condensate, just like the Schwinger pair production results in the discharge of the background electric field. This depletion can be easily quantified using the scattering rate of the constituents
\beq
\frac{\dot{N}}{N}=H\left( \epsilon-\frac{1}{N} \right)\,.
\label{depletion}
\eeq
Here, the first term on the right hand side is a classical increase of the number of constituent gravitons as a result of the slow-roll, while the second term represents the quantum depletion. This relation simply tells us that quantum mechanically the condensate loses approximately one constituent graviton in every Hubble time. This immediately suggests that the semi-classical description is valid as long as the first term dominates over the second one in \eqref{depletion}. In other words we have the following consistency bound
\beq
\epsilon > \frac{1}{N}\,.
\label{consb}
\eeq
As one might have expected, this coincides with the threshold of self-reproduction for the $m^2\f^2$-model. This inequality can be rewritten as the upper bound on number of e-folds
\beq
\mathcal{N}_{\rm e}<N\,.
\eeq
The physical meaning of these bounds is very simple. If we begin inflation with $\epsilon<1/N$, then the entire condensate will be depleted before the end of inflation\footnote{We would like to point out that, in the presence of the large number of light particles (with mass less than the Hubble scale), the consistency bound \eqref{consb} would read
\beq
\epsilon > \frac{n_{\rm species}}{N}\,.\nonumber
\eeq
This would, obviously, lower the upper bound on number of e-folds down to $N/n_{\rm species}$.
}.

We would like to emphasize that the depletion channel, identified in the current work as the relevant one, is independent of the occupation number of the constituent inflatons. Therefore, it should not be surprising that the obtained bound on the lifetime of the inflationary background is identical to the quantum break-time of the de Sitter space found in \cite{Dvali:2014gua}.

In this section we have considered two sources for changing number of constituent gravitons: classical slow-roll and the depletion through the scattering of the constituents. However, there is a third source not included in \eqref{depletion}, namely the density perturbations. In particular, the graviton condensate is a dressing field for the scalar field configuration, dictated by the quantum constraint equations and correspondingly by the consistency with the underling diffeomorphism invariance. Hence, scalar field fluctuations, which originate from the uncertainty in the position of the constituent inflatons, will necessarily result in the fluctuation of the number of constituent gravitons.

To be more specific, we have argued in the previous section that the typical density perturbation can be estimated as \eqref{UPperturb}. This results in the following change in the Hubble scale
\beq
\delta H =\frac{mH}{\mpl}\,,
\eeq
which results in the following change in the number of gravitons within the Hubble patch
\beq
\delta N =\frac{\mpl m}{H^2}\,.
\eeq
The depletion equation \eqref{depletion} is valid only if $\delta N \ll 1$. It is straightforward to show that this is indeed so in the self-reproductive regime $\epsilon \ll 1/N$.

\section{Summary}

We have argued that the homogeneous classical scalar field of the inflationary background corresponds to the Bose-Einstein condensate of the off-shell inflaton degrees of freedom with the mass gap of the order of $m^2/H$. Furthermore, we have shown that the physics behind the origin of the density perturbations depends on the curvature scale. In the slow-roll regime, the perturbations originate from the quantum uncertainty in the position of the constituent inflatons. While, in the regime of eternal inflation, they originate from the annihilation of the constituent gravitons. The latter process is also responsible for the production of the primordial gravitational waves.

The annihilation of the constituent gravitons causes the gradual depletion of the graviton condensate, giving rise to the upper bound on the lifetime of the quasi-de Sitter space. Namely, as it was first suggested in \cite{Dvali:2014gua}, the Hubble patch of the de Siiter space with the curvature scale $H$ has $N=\mpl^2/H^2$ number of longitudinal gravitons in the zero momentum state. Thus, if the condensate loses approximately one graviton in the Hubble time it will take $N$ e-folds to deplete the entire reservoir. In the current work, we have shown that the situation of the slow-roll inflation is very similar to the case of the de Sitter space. If the Hubble patch with $N$ longitudinal gravitons had started deep in the self-reproductive regime, then after spending of order $N$ e-folds in this regime it would have ceased to be described by a semi-classical geometry. 

\vskip 20pt

{\bf Acknowledgements:} I am especially grateful to Gia Dvali for illuminating discussions and comments. I would also like to thank Justin Khoury, Valery Rubakov and Herman Verlinde for useful discussions, and Jean-Luc Lehners for useful comments on the manuscript. This work is supported by US Department of Energy grant DE-SC0007968.

\section*{Appendix: Fluctuations in the gas of indistinguishable particles}
\renewcommand{\theequation}{A-\Roman{equation}}
\setcounter{equation}{0} 

Considering a large box of size $V$ and total number of particles $N$, it is easy to show that the probability distribution for the number of particles $n$ inside the smaller volume $v\ll V$ is given by the Poisson's distribution with
\beq
\la n \ra&=&N \frac{v}{V}\,,\\
\la \delta n^2 \ra&=& \la n \ra\,.
\eeq
Strictly speaking, this is true for the distinguishable particles. For the indistinguishable ones, on the other hand, quantum fluctuations tend to be much more pronounced. In fact, the result is sensitive to quantum state the particles are in. If the state is a direct product of one-particle-states or a coherent one, then one gets the same result as for the distinguishable ones. However, if the system is believed to be in equilibrium and hence all physically different configurations are assigned the same probability, then the situation changes drastically.

To discuss the latter case in detail, let us consider the above-mentioned large box to have been split into $r\equiv V/v$ number of equal size cells. If we label the states of the system in terms of the number of particles in each cell, then every possible state can be written as
\beq
(n_1,n_2,\ldots,n_r)\,,
\eeq
with $\sum n_i=N$. Moreover, in case of the uniformly distributed non-interacting particles, all these states must have equal probabilities. The total number of such states is
\beq
\frac{(r+N-1)!}{(r-1)!N!}\,.
\eeq
Now, we would like to find a probability of having $n$ particles in one of the cells, e.g. $n_1=n$. This can be done by dividing the number of distinguishable states with $n_1=n$ by the total number of distinguishable states
\beq
P(n_1=n)=\frac{(r+N-n-2)!}{(r-2)!(N-n)!}\times\frac{(r-1)!N!}{(r+N-1)!}\,.
\eeq
As a check for the proper normalization we have
\beq
\sum_{n=0}^{n=N}P(n_1=n)=1\,.
\eeq
We can also find the expectation value of $n$, as
\beq
\la n \ra =\sum_{n=0}^{n=N}n P(n_1=n)=\frac{N}{r}\,.
\eeq
Which is an intuitive result, considering the uniform distribution. The amplitude of the number fluctuation can be computed in a similar way
\beq
\la \delta n^2 \ra=\la n^2 \ra - \la n \ra^2=\sum_{n=0}^{n=N}n^2 P(n_1=n)-\la n \ra^2\,.
\eeq
After a little bit of algebra, we obtain
\beq
\la \delta n^2 \ra=\la n\ra \big(1+\la n\ra\big)\frac{r-1}{r+1}\,.
\eeq
This may look somewhat puzzling, since in case of $\la n\ra\gg1$ and $r\gg 1$ we have
\beq
\la \delta n^2 \ra\simeq\la n\ra^2\,.
\eeq
Which implies order one number fluctuation.

\end{document}